\documentclass[superscriptaddress,showpacs,nofootinbib,showkeys,10pt]{article}
\usepackage[titletoc,title,toc]{appendix}
\usepackage{graphicx}  
\usepackage{amsmath}   
\usepackage[compress]{cite}
\usepackage{amssymb}   
\usepackage{bm} 
\usepackage{dcolumn}
\usepackage{color}
\usepackage{mathrsfs}
\usepackage{amsfonts}
\usepackage{longtable}
\usepackage{varioref}
\RequirePackage[colorlinks,citecolor=blue,urlcolor=magenta,linkcolor=blue]{hyperref}
\def\ntabla{\bar{\nabla}}
\labelformat{section}{Section #1} 
\labelformat{subsection}{Section #1} 
\labelformat{subsubsection}{Section #1}
\labelformat{subsubsubsection}{Section #1}
\labelformat{equation}{Eq.~(#1)} 
\labelformat{figure}{Fig.~#1} 
\labelformat{subfigure}{Fig.~\thefigure#1} 
\labelformat{table}{Table~#1} 
\labelformat{appendix}{Appendix #1}
\title{Late-time acceleration driven by shift-symmetric Galileon in the presence of Torsion}

\author{Rabin Banerjee\footnote{rabin@bose.res.in}$~^{1}$, 
Sumanta Chakraborty\footnote{sumantac.physics@gmail.com}$~^{2}$ and 
Pradip Mukherjee\footnote{mukhpradip@gmail.com}$~^{3}$\\
{\small{$^{1}$S. N. Bose National Centre for Basic Sciences}}\\
{\small{JD Block, Sector III, Salt Lake City, Kolkata 700098, India}}\\
{\small{$^{2}$Department of Theoretical Physics}}\\
{\small{Indian Association for the Cultivation of Science, Kolkata 700032, India}}\\
{\small{$^{3}$Department of Physics, Barasat Government College}}\\
{\small{10, KNC Road, Barasat, Kolkata 700124, India}}}

\begin{document}
\maketitle
\begin{abstract}
A shift-symmetric Galileon model in presence of spacetime torsion has been constructed for the first time. This has been realized by localizing (or, gauging) the Galileon symmetry in flat spacetime in an appropriate manner. We have applied the above model to study the evolution of the universe at a cosmological scale. Interestingly, for a wide class of torsional structures we have shown that the model leads to late time cosmic acceleration. Furthermore, as torsion vanishes, our model reproduces the standard results.   
\end{abstract}
\paragraph*{\bf Introduction} ---~~ Late time acceleration of the universe is one of the most profound as well as intriguing discoveries of this century \cite{Smoot:1992td,Perlmutter:1998np,Riess:1998cb,Komatsu:2010fb,Kowalski:2008ez,
Copeland:2006wr,Ade:2013zuv}. This suggests that despite gravity being attractive, at large enough length scales there is a repulsive effect causing an accelerated expansion of our universe. This has caused a huge debate over the scientific community, whether this implies existence of some exotic matter, normally referred to as dark energy or it may indicate breakdown of Einstein's theory at large scales \cite{Carroll:2000fy,Peebles:2002gy,Padmanabhan:2002ji}. There have been numerous candidates so far which attempt to explain the observed accelerated expansion of the universe by either modifying the Einstein-Hilbert action by incorporating higher curvature terms or by introducing additional matter fields, in the form of a perfect fluid, with equation of state parameter $\omega <-(1/3)$. However most of these models introduced to address the late time acceleration of the universe, do not follow from basic principles. Naturally, if one can obtain such a model from basic principles, yet can explain late time acceleration, it is bound to be a very attractive choice. The model presented in this work is an attempt in this direction.

The Galileon Lagrangian that originated from the Dvali-Gabadadze-Porrati (DGP) model \cite{Dvali:2000hr,Luty:2003vm} has two special properties --- (a) in spite of being a higher derivative theory it is guaranteed to produce second order field equations and (b) it is endowed with  the shift symmetry ($\textrm{scalar}\rightarrow \textrm{scalar}+b_{\mu}x^{\mu}+c$, where $b_{\mu}$ is a constant vector and $c$ is a constant number). These twin features make the model ghost free along with some special renormalization properties \cite{Nicolis:2008in,Deffayet:2013lga,Goon:2016ihr}. These properties suggest that the Galileon model, appropriately tailored for curved spacetime could be a viable candidate, which can explain the late time acceleration of the universe without invoking exotic matters \cite{Chow:2009fm,Gannouji:2010au,Pirtskhalava:2015nla,Gleyzes:2014qga}. Unfortunately, it is known that the Galileon model runs into some difficulty when gravity is introduced. This is due to the impossibility of coupling the Galileon field with gravity, while retaining the complete shift symmetry \cite{Deffayet:2009wt}. The reason is not hard to understand. A constant vector representing the gradient shift of the flat spacetime Galileon model is incompatible with curved spacetime.

The construction of such a shift symmetric Galileon model in curved spacetime hinges on the fact that, symmetries on flat spacetime hold locally in the curved spacetime. Taking a cue from the above, the powerful method of localizing both relativistic symmetry \cite{Utiyama:1956sy,Kibble:1961ba} as well as non-relativistic symmetry \cite{Banerjee:2014pya,Banerjee:2015rca} was used to obtain the Galileon Lagrangian which respects the full shift symmetry \emph{locally} \cite{Banerjee:2017qdl}. Subsequently, a metric formulation was developed in \cite{Banerjee:2017jyb}. Interestingly, the covariant Galileon models formulated so far were obtained in the context of Riemannian manifold and do \emph{not} involve spacetime torsion. Since, a priori there is no reason for the spacetime torsion to be vanishing it is instructive to ask what happens to the Galileon Lagrangian when the spacetime torsion is non-zero. Note that in \cite{Banerjee:2017qdl} the spacetime torsion is assumed to be vanishing, which is just a 
choice and we always have the option of including torsion in the model as the main theory is formulated in the vielbein approach. In particular, in view of \cite{Ivanov:2016xjm}, where spacetime torsion was introduced in a different context to explain the late time acceleration of the universe, inclusion of torsion in the Galileon model may lead to an accelerated expansion of the universe. This is what we would like to discuss here. 
 
In this letter we will explicitly demonstrate that it is indeed possible to construct a covariant shift symmetric Galileon model to explain the late time cosmic acceleration of our universe, but \emph{only} when torsion is included. The novelty of our studies are three fold --- (a) This is the first venture in Galileon cosmology which includes the effect of torsion, (b) Eventually we will see that the Galileon field itself can act as a source of torsion and finally (c) It does not depend on any ad hoc element but follows naturally from localizing the symmetries. The Galileon model presented here contains the full Galileon shift symmetry \emph{locally}, reproducing the standard Galileon model in the flat limit.
 
The letter is organized as follows: we will first introduce spacetime torsion in the Galileon Lagrangian and discuss the essential changes brought into the matter sector. Modifications of the gravitational field equations due to inclusion of spacetime torsion will also be addressed subsequently. Finally we will demonstrate how a non-trivial configuration of the torsion tensor can lead to accelerating solutions of the universe. We will set the fundamental constants $c=1=G$, the Greek indices will denote spacetime coordinates, while the roman indices will stand for coordinates of the local inertial frame. Furthermore, throughout the letter we will denote all the quantities involving torsion with a ``bar" on the respective quantity. 
\paragraph*{\bf Covariant Galileon Model with torsion} ---~~ We will construct the Galileon Lagrangian in presence of spacetime torsion by adopting the techniques of localization of shift-symmetry along  with the Poincar\'e symmetries in flat spacetime. Such localization yields the following action for the Galileon field $\pi$ in the local inertial frame, irrespective of the presence of spacetime torsion \cite{Banerjee:2017qdl} 
\begin{align}\label{Local_Galileon_Action}
\bar{\mathcal{A}}_{\pi}=\int d^{4}x~\frac{1}{\Sigma}&\Big[-\frac{1}{2}\eta ^{ab}\tilde{\mathcal{D}}_{a}\pi\tilde{\mathcal{D}}_{b}\pi 
\nonumber
\\
&+\alpha \left(\eta ^{ab}\tilde{\mathcal{D}}_{a}\pi\tilde{\mathcal{D}}_{b}\pi \right)\left(\eta ^{cd}\mathcal{D}_{c}\tilde{\mathcal{D}}_{d}\pi\right)\Big]~.
\end{align}
Here the covariant derivative operator $\tilde{D}_{a}$ is defined by its action on the Galileon field $\pi$ as: $\tilde{\mathcal{D}}_{a}\pi \equiv \Sigma ^{\mu}_{a}D_{\mu}\pi \equiv \Sigma ^{\mu}_{a}(\partial _{\mu}\pi +F_{\mu})$, where $F_{\mu}$ is the gauge field originated from localization of shift-symmetry in flat spacetime. The object $\Sigma ^{\mu}_{a}$ is the tetrad field satisfying $\eta ^{ab}\Sigma ^{\mu}_{a}\Sigma ^{\nu}_{b}=g^{\mu \nu}$, where $\eta _{ab}=\textrm{diag}(-1,1,1,1)$ is the Minkowski metric. Note that the above Lagrangian is not the most general Galileon Lagrangian in four spacetime dimensions, as in principle there can be two more terms present in \ref{Local_Galileon_Action} \cite{Deffayet:2013lga,Gleyzes:2014qga}. However, as we will demonstrate, keeping the first non-trivial term (i.e., the term with coefficient $\alpha$) alone will serve our purpose. Thus in what follows we will concentrate on the simplest non-trivial Galileon model. 

It is possible to write down the above action in a general curved spacetime background by using the tetrad field $\Sigma ^{a}_{\mu}$ and transforming from local inertial coordinates $\{x^{a}\}$ to general coordinate system $\{y^{\mu}\}$. Performing the indicated coordinate transformation we obtain the following action for the Galileon field in curved spacetime in presence  of torsion
\begin{align}\label{Final_Action}
\bar{\mathcal{A}}_{\pi}&=\int d^{4}x~\sqrt{-g}\Big[-\frac{1}{2}g^{\alpha \beta}\left(\partial _{\alpha}\pi+F_{\alpha}\right)\left(\partial _{\beta}\pi+F_{\beta}\right)
\nonumber
\\
&\hspace{0.5cm}+\alpha \Big\{g^{\alpha \beta}\left(\partial _{\alpha}\pi+F_{\alpha}\right)\left(\partial _{\beta}\pi+F_{\beta}\right)\Big\}
\nonumber
\\
&\hspace{1cm}\times\Big\{\square \pi +\partial _{\mu}F^{\mu}-T_{\mu}\left( \partial ^{\mu}\pi+F^{\mu}\right)\Big\}\Big]~.
\end{align}
Here the term inside square bracket corresponds to the Lagrangian density for the Galileon field $\bar{L}_{\pi}$. In order to arrive at the terms in the last line of the above action we have used the following property of the tetrad field: $\Sigma ^{\mu}_{c}D_{\mu}\Sigma ^{\nu}_{d}=-\Sigma ^{\mu}_{c}\Sigma ^{\rho}_{d}\bar{\Gamma}^{\nu}_{~\mu \rho}$. Here $\bar{\Gamma}^{\alpha}_{~\mu \nu}$ is the connection associated with covariant derivative in presence of spacetime torsion $T^{\alpha}_{~\mu \nu}$, defined as $T^{\alpha}_{~\mu \nu} \equiv \bar{\Gamma}^{\alpha}_{~\mu \nu}-\bar{\Gamma}^{\alpha}_{~\nu \mu}$. As evident, the torsion field $T^{\alpha}_{~\mu \nu}$ is antisymmetric in the $(\mu ,\nu)$ indices. Moreover the quantity $T_{\mu}$ appearing in \ref{Final_Action} is the trace of the torsion tensor and is defined as, $T_{\mu}=T^{\alpha}_{~\mu \alpha}$. 

For completeness, let us spell out the transformation properties of the Galileon field $\pi$ and the gauge field $F_{\mu}$. Under Galileon transformation, we have $\delta \pi=c+b_{\nu}x^{\nu}$ and $\delta F_{\mu}=-\partial _{\mu}c-x^{\nu}\partial _{\mu}b_{\nu}$, such that the action remains invariant modulo total derivative terms. Note that due to the presence of the gauge field $F_{\mu}$, the full Galileon symmetry is respected by the Lagrangian even in curved spacetime. This being the prime reason for appearance of the gauge field in the Galileon action. Note that the presence of the gauge field $F_{\mu}$ can also be ascribed to have originated from a St\"{u}ecklberg transformation. Thus there exist a possibility to add additional kinetic terms for the gauge field, however our concern is essentially about the Galileon Lagrangian \emph{alone} and hence such terms will not be considered in this work.

Further, the equation satisfied by the gauge field $F^{\mu}$ also gets modified in presence of spacetime torsion and takes the following form:
\begin{align}\label{Eq_For_Fmu}
\ntabla _{\mu}F_{\nu}-\ntabla _{\nu}F_{\mu}
=\partial _{\mu}F_{\nu}-\partial _{\nu}F_{\mu}-T^{\rho}_{\mu \nu}F_{\rho}=0~,
\end{align}
where $\ntabla$ is the covariant derivative which incorporates the effect of torsion as well. Therefore, in the general context in order to obtain the gauge field one needs to solve \ref{Eq_For_Fmu} in an explicit manner. Thus \ref{Eq_For_Fmu} provides the evolution equation for the gauge field $F_{\mu}$, which can be solved if time evolution of torsion is known. Hence the time evolution of the gauge field depends on its interaction with the torsional degrees of freedom. Note that we did not have to add any kinetic term for the gauge field to arrive at the above evolution equation.

Before pursuing the general case, it will be worthwhile to observe the corresponding situation in absence of spacetime torsion. In this case the unique solution for \ref{Eq_For_Fmu} becomes $F_{\mu}=\partial _{\mu}\Phi$, for some scalar function $\Phi$. This enables one to write down the Lagrangian for the curved spacetime Galileon field in absence of spacetime torsion as, $L_{\pi}=-(1/2)\partial _{\mu}\pi'\partial ^{\mu}\pi '+\alpha (\partial _{\mu}\pi'\partial ^{\mu}\pi ')\square \pi'$, where $\pi'=\pi+\Phi$. As evident from the structure, the above Lagrangian is essentially the covariant Galileon Lagrangian incorporating the kinetic term and first non-trivial correction to the Galileon Lagrangian, known as the $L_{3}$ term. Therefore our action as presented in \ref{Final_Action} indeed reduces to the correct action for curved spacetime Galileon field in the absence of torsion. From the above analysis it is clear that in presence of torsion the solution $F_{\mu}=\partial _{\mu}\Phi$ does not satisfy \ref{Eq_For_Fmu}, which leads to additional new structures in the theory. 

Let us first ensure that the modified action for the Galileon field, as presented in \ref{Final_Action} satisfies a second order field equation in $\pi$. This can be achieved by varying \ref{Final_Action} with respect to  the Galileon field. Since the torsional part has already been separated out in \ref{Final_Action}, one can use the fact that $\nabla _{\mu}\nabla _{\nu}\pi=\nabla _{\nu}\nabla _{\mu}\pi$ and the commutator of covariant derivative can be written in terms of Riemann tensor. Thus it is straightforward to explicitly verify that the field equation for the Galileon field does not contain more than second derivatives of $\pi$ and single derivative of the gauge field. Therefore the action for the Galileon field indeed evades the Ostrogradsky instability and therefore is free from any ghost modes. 

Having ensured the stability of the theory, we now consider the energy-momentum tensor for the Galileon field which we will need in order to write down the gravitational field equations. To derive the energy-momentum tensor starting from the Galileon action $\bar{\mathcal{A}}_{\pi}$, one should vary the action with respect to the metric tensor $g^{\mu \nu}$, yielding, 
\begin{align}\label{EMT_Gal}
\bar{T}^{(\pi)}_{\mu \nu}&\equiv -\frac{2}{\sqrt{-g}}\frac{\delta \bar{\mathcal{A}}_{\pi}}{\delta g^{\mu \nu}}
=\left(\nabla _{\mu}\pi +F_{\mu}\right)\left(\nabla _{\nu}\pi+F_{\nu}\right)
-\frac{1}{2}g_{\mu \nu}\Big\{g^{\alpha \beta}\left(\partial _{\alpha}\pi+F_{\alpha}\right)
\left(\partial _{\beta}\pi+F_{\beta}\right)\Big\}
\nonumber
\\
&+\alpha \Bigg[\nabla _{\mu}\Big\{g^{\alpha \beta} \left(\partial_{\alpha}\pi+F_{\alpha}\right)
\left(\partial _{\beta}\pi+F_{\beta}\right)\Big\} 
\left(\nabla _{\nu}\pi+F_{\nu}\right)
+\nabla _{\nu}\Big\{g^{\alpha \beta}\left(\partial _{\alpha}\pi+F_{\alpha}\right)
\left(\partial _{\beta}\pi+F_{\beta}\right)\Big\}
\left(\nabla _{\mu}\pi+F_{\mu}\right)
\nonumber
\\
&\hspace{1cm}-2\Big(\square \pi+\nabla _{\alpha}F^{\alpha}\Big)\left(\nabla _{\mu}\pi+F_{\mu}\right)
\left(\nabla _{\nu}+F_{\nu}\right)
-g_{\mu \nu}\nabla _{\rho}\Big\{g^{\alpha \beta}\left(\partial _{\alpha}\pi+F_{\alpha}\right)
\left(\partial _{\beta}\pi+F_{\beta}\right)\Big\}
\left(\nabla ^{\rho}\pi+F^{\rho}\right)\Bigg]
\nonumber
\\
&+\alpha \Bigg[\left(\partial _{\alpha}\pi +F_{\alpha}\right)\left(\partial ^{\alpha}\pi+F^{\alpha}\right)
\Big\{T_{\mu}\left(\partial _{\nu}\pi+F_{\nu}\right)
+T_{\nu}\left(\partial _{\mu}\pi+F_{\mu}\right)\Big\}
\nonumber
\\
&\hspace{0.5cm}+2\Big\{T_{\alpha}\left(\partial ^{\alpha}\pi +F^{\alpha}\right)\Big\}
\left(\partial _{\mu}\pi +F_{\mu}\right)\left(\partial _{\nu}\pi+F_{\nu}\right)-g_{\mu \nu}\Big\{T_{\alpha}\left(\partial ^{\alpha}\pi+F^{\alpha}\right)\left(\partial _{\beta}\pi+F_{\beta}\right)
\left(\partial ^{\beta}\pi+F^{\beta}\right)\Big\}\Bigg]~.
\end{align}
The terms present in the first three lines of the above energy momentum tensor is what one would have obtained if spacetime torsion was absent and with the choice $F_{\mu}=\partial _{\mu}\psi$ it reduces to the energy momentum tensor derived for covariant Galileon models described in the literature \cite{Deffayet:2013lga,Banerjee:2017jyb}. Note that the field equation for the Galileon field $\pi$ can also be obtained from the covariant conservation of the stress-energy tensor, i.e., $\nabla _{\mu}\bar{T}^{\mu~(\pi)}_{\nu}=0$. 

The above energy momentum tensor associated with the Galileon field $\bar{T}_{\mu \nu}^{(\pi)}$ should sit on the right hand side of the Einstein's equations, which will inherit further corrections due to the presence of spacetime torsion, affecting the gravitational Lagrangian as well. Finally assuming the background geometry to be homogeneous and isotropic, representing our universe at large scales, one can immediately write down the Friedmann equations, from which the late time acceleration of our universe will follow. 
\paragraph*{\bf Gravity in presence of torsion}---~~ We have discussed the effect of spacetime torsion on the matter sector, namely the Galileon field. As evident, there will also be modifications in the geometrical sector as well due to presence of spacetime torsion, which we will consider now. Due to presence of torsion the definition of Riemann tensor through commutator of covariant derivative itself will be modified, which in turn will alter the expressions for Ricci tensor and Ricci scalar, thereby the Einstein tensor will inherit corrections. In particular, the commutator of covariant derivative in presence of spacetime torsion reads,
\begin{align}
\left[\bar{\nabla}_{\mu},\bar{\nabla}_{\nu}\right]A_{\alpha}=\bar{R}_{\mu \nu \alpha}^{~~~~\rho}A_{\rho}-T^{\rho}_{~\mu \nu}\bar{\nabla}_{\rho}A_{\alpha}~.
\end{align}
From which it is straightforward to compute the Riemann tensor, which can be divided into two parts, one independent of spacetime torsion and the other dependent on spacetime torsion. Therefore given the Riemann tensor one may compute the Ricci tensor as well as the Ricci scalar subsequently and divide them into a part originating from the metric degrees of freedom and another one from the spacetime torsion. This results into, $\bar{R}=R+(K^{\alpha}_{~\alpha \rho}K^{\rho \beta}_{~~~\beta}-K^{\rho}_{~\alpha \mu}K^{\alpha \mu}_{~~~\rho})$. Here we have neglected total divergences and have defined the contorsion tensor $K^{\mu}_{~\alpha \beta}$ appearing in the above expression as, $K^{\mu}_{~\alpha \beta}\equiv (1/2)(T^{\mu}_{~\alpha \beta}-T_{\alpha \beta}^{~~~\mu}-T_{\beta \alpha}^{~~~\mu})$. Starting from the expressions for Ricci tensor and Ricci scalar one can also construct a similar relationship between the Einstein tensor with and without torsion. Finally to derive the gravitational field equations in presence of torsion it is necessary to vary both the gravitational action involving $\bar{R}$, as well as the action $\bar{\mathcal{A}}_{\pi}$ for the Galileon field with respect to the metric. The variation of the action for Galileon field will lead to the energy momentum tensor $\bar{T}^{(\pi)}_{\mu \nu}$, while variation of the gravitational action will result into the Einstein tensor along with additional contributions originating from the torsion field. Therefore, the field equations for $(\textrm{gravity+Galileon~Field+Torsion})$ 
system become
\begin{align}\label{Grav_Field_Eq}
G_{\mu \nu}&=T_{\mu}T_{\nu}-K^{\alpha}_{~\beta \mu}K^{\beta}_{~\alpha \nu}-\frac{1}{2}g_{\mu \nu}\left(T^{\rho}T_{\rho}-K_{\alpha \mu \rho}K^{\mu \alpha \rho}\right)
\nonumber
\\
&+8\pi G \bar{T}^{(\pi)}_{\mu \nu}~,
\end{align}
where $\bar{T}^{(\pi)}_{\mu \nu}$ has been presented in \ref{EMT_Gal}. The above presents the dynamical equation for the metric $g_{\mu \nu}$. However the torsion tensor still remains undetermined. To derive a source for the torsion tensor we can follow an identical path and one should vary the total Lagrangian density $\bar{R}+\bar{L}_{\pi}$ with respect to the contorsion tensor $K^{\alpha}_{~\mu \nu}$. Variation of the gravitational Lagrangian with respect to contorsion tensor leads to, $(\delta \bar{R}/\delta K^{\mu}_{~\sigma \nu})\equiv S^{\sigma~\nu}_{~\mu}=T^{\sigma~\nu}_{~\mu}+T^{\nu}\delta ^{\sigma}_{\mu}-T_{\mu}g^{\sigma \nu}$. Along identical lines the corresponding source term originating from the Galileon Lagrangian is defined as $(\delta \bar{L}_{\rm \pi}/\delta K^{\mu}_{~\alpha \beta})\equiv -\bar{\tau} ^{\alpha~\beta~(\pi)}_{~\mu}$. Given the Galileon Lagrangian $\bar{L}_{\pi}$ the tensor $\bar{\tau}^{\mu~(\pi)}_{~\alpha \beta}$ becomes, 
\begin{align}\label{taubar_Gal}
\bar{\tau}^{\mu~(\pi)}_{~\alpha \beta}&=\frac{1}{2}\Big[\delta ^{\mu}_{\beta}\left(\partial _{\alpha}\pi+F_{\alpha}\right)
\left\{\left(\partial _{\rho}\pi+F_{\rho}\right)\left(\partial ^{\rho}\pi+F^{\rho}\right) \right\}
\nonumber
\\
&-\delta ^{\mu}_{\alpha}\left(\partial _{\beta}\pi+F_{\beta}\right)
\left\{\left(\partial _{\rho}\pi+F_{\rho}\right)\left(\partial ^{\rho}\pi+F^{\rho}\right) \right\}\Big]~.
\end{align}
Thus Galileon field can act as the source of spacetime torsion! Therefore setting the total variation of the matter plus gravitational action to be vanishing for arbitrary variation of the contorsion tensor, we obtain $S^{\mu}_{~\alpha \beta}=16\pi G ~\bar{\tau}^{\mu~(\pi)}_{~\alpha \beta}$. Since the tensor $S^{\mu}_{~\alpha \beta}$ can be expressed in terms of either the torsion tensor $T^{\mu}_{~\alpha \beta}$ or the contorsion tensor $K^{\mu}_{~\alpha \beta}$, one can relate the torsion tensor and the contortion tensor with $\bar{\tau}^{\alpha~(\pi)}_{~\mu \nu}$ using the above result. These expressions can in turn be substituted in the gravitational field equations, such that the right hand side of the Einstein's equations depend not only on the Galileon field $\pi$ and gauge field $F_{\mu}$, but also on various quadratics of $\bar{\tau}^{\mu~(\pi)}_{~\alpha \beta}$ and its trace. This provides the general prescription to derive the gravitational field equations in presence of spacetime torsion and 
Galileon field. In the next section we will apply the formalism developed here to homogeneous and isotropic spacetime, to understand the dynamics of our universe at a large scale.
\paragraph*{\bf Cosmological Implications}---~~ As an application of the formalism presented above we will discuss cosmological implications of the Galileon field in presence of spacetime torsion in this section. This requires one to assume that the spacetime is homogeneous and isotropic and therefore is described solely by the scale factor $a(t)$, which is a function of time alone. In what follows we will further assume the spatial curvature of the spacetime to be vanishing in conformity with recent observations \cite{Ade:2013zuv}.

The homogeneity and isotropy of the spacetime demands $\pi=\pi(t)$, i.e., the Galileon field $\pi$ is an explicit function of time alone. Moreover, the gauge field $F_{\mu}$ must take the following form, $F_{\mu}=\left\{F(t),f(t),f(t),f(t)\right\}$. Here $F(t)=\partial _{t}\Phi$ and $f(t)$ are arbitrary functions of time. Note that this is the unique structure of the gauge field that preserves the isotropy and homogeneity of the spacetime. Moreover a similar consideration applies to spacetime torsion as well, since the trace $T^{\alpha}_{~\mu \alpha}$ of the spacetime torsion appears in the Galileon Lagrangian, it is legitimate to take $T_{\mu}=(T,\tau,\tau,\tau)$. Here also $T=T(t)$ and $\tau=\tau(t)$ are functions of time alone. Therefore the differential equation satisfied by the gauge field $F_{\mu}$ presented in \ref{Eq_For_Fmu} takes the following form: $\partial _{t}f+(\tau/3)F-(T/3)f=0$. Given the above, one can immediately obtain various components of the energy-momentum tensor presented in \ref{EMT_Gal}. Further the tensor $\bar{\tau} ^{\mu}_{~\alpha \beta}$, introduced in \ref{taubar_Gal}, can be written down in terms of two independent quantities --- (a) $\bar{\tau}'=(\dot{\pi}'/2)\{-\dot{\pi}'^{2}+(3f^{2}/a^{2})\}$ and (b) $\bar{\tau}''=(f/2)\{-\dot{\pi}'^{2}+(3f^{2}/a^{2})\}$, where $\pi'=\pi+\Phi$. Therefore using the results for the gauge field, torsion tensor and $\bar{\tau}^{\mu}_{~\alpha \beta}$ one can determine the right hand side of \ref{Grav_Field_Eq}. In particular, it turns out that the diagonal entries, i.e., $T^{t}_{t}$, $T^{x}_{x}$, $T^{y}_{y}$ and $T^{z}_{z}$ are non-zero and furthermore $T^{x}_{x}=T^{y}_{y}=T^{z}_{z}$. Hence it is possible to identify the energy density $\rho$ and pressure $p$ associated with the Galileon field in presence of torsion as, 
\begin{align}
\rho&=\frac{\dot{\pi}'^{2}}{2}+6\alpha H\dot{\pi}'^{3}
+\frac{3f^{2}}{2a^{2}}\Big[1-4\alpha H\dot{\pi}'+4\alpha \frac{\dot{f}}{f}\dot{\pi}'\Big]
+\alpha \Big[-3T\dot{\pi}'^{3}+3\frac{tf}{a^{2}}\dot{\pi}'^{2}
\nonumber
\\
&+3T\dot{\pi}'\frac{f^{2}}{a^{2}}+9t\frac{f^{3}}{a^{4}}\Big]-8\pi G \left[3\tau'^{2}+\frac{9}{a^{2}}\tau''^{2}\right]~,
\label{Eq_Gal_rho}
\\
p&=\frac{\dot{\pi}'^{2}}{2}-2\alpha \dot{\pi}'^{2}\ddot{\pi}'
-\frac{f^{2}}{2a^{2}}\Big[1-4\alpha \ddot{\pi}'-12\alpha \dot{\pi'}\frac{\dot{f}}{f}\Big]
+\alpha \Big[T\dot{\pi}'\frac{f^{2}}{a^{2}}+3t\frac{f^{3}}{a^{4}}+\dot{\pi}'^{2}\frac{tf}{a^{2}}
\nonumber
\\
&-T\dot{\pi}'^{3}\Big]+8\pi G \left[-\frac{3}{a^{2}}\tau''^{2}+3\tau'^{2}\right]~.
\label{Eq_Gal_p}
\end{align}
The other components of the energy-momentum tensor sitting on the right hand side of gravitational field equations must vanish. This in turn provides the two independent components of the torsion tensor $T$ and $\tau$ in terms of $f$ and $\pi'$ respectively. 

\begin{figure}[t!]
\centering
\includegraphics[scale=1]{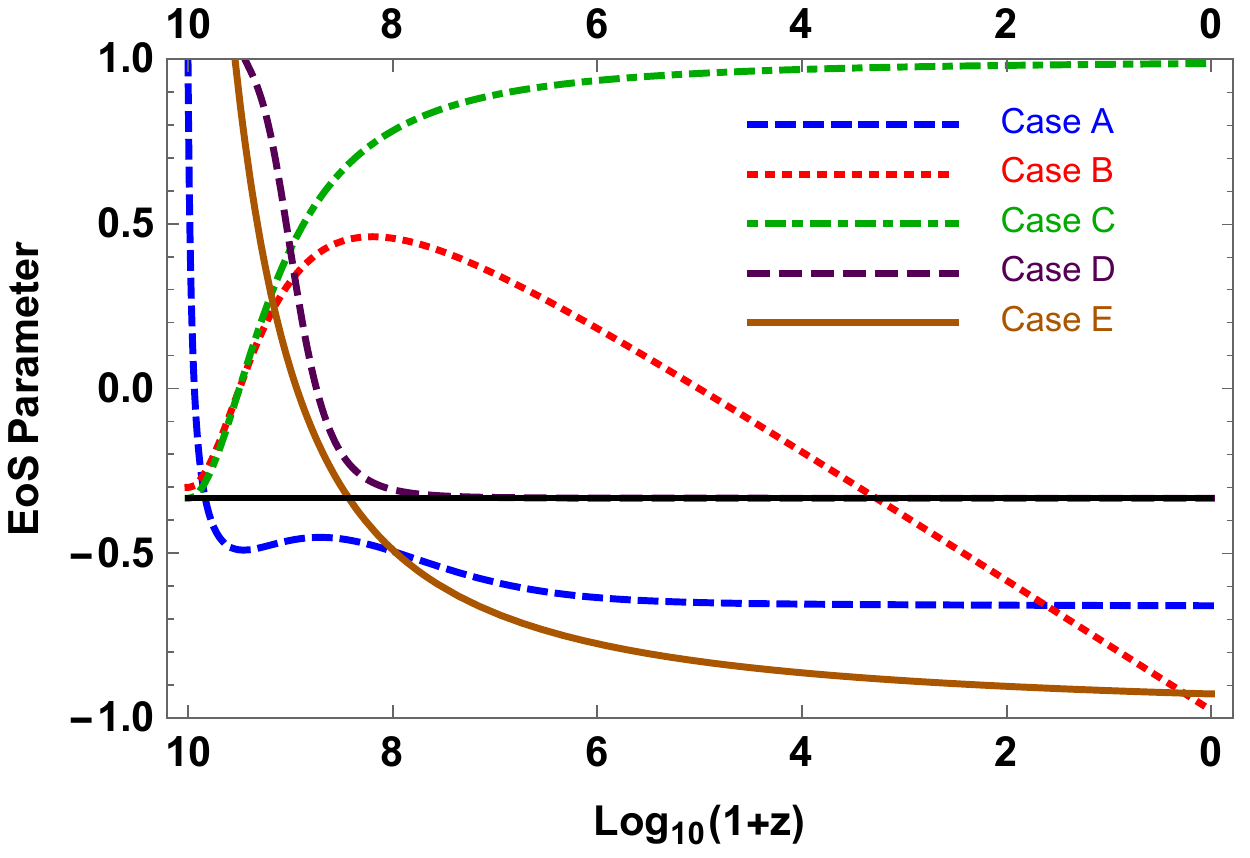}~~
\caption{The above figure demonstrates the evolution of the Equation of State (EoS) parameter $\omega \equiv p/\rho$ for the Galileon field coupled to spacetime torsion. The black line stands for the $\omega=-1/3$, such that if any curve for $\omega$ are below this line, it will lead to accelerating expansion. However as current observational estimates suggest \cite{Ade:2013zuv}, $\omega _{0}=-1$ is more favoured as late time cosmic acceleration is concerned. As evident there are two cases --- (i) Case B: In which fields except the gauge field is evolving as power law, while gauge field is evolving exponentially; (iii) Case E: In which the power law behaviour of the fields are different and $\alpha$ is small, we have late time cosmic acceleration in our model in the present epoch. See text for more discussions.}
\label{Fig_EoS}
\end{figure}

Therefore, one must have a non-trivial choice for the gauge field with both $F$ and $f$ non-zero. This in turn will require both $T$ and $\tau$ to be non-vanishing. Therefore in this general context we will have the evolution of the universe to be driven by the energy density $\rho$ and pressure $p$ having contributions from the Galileon field $\pi$, the gauge field $F_{\mu}$ and the torsion tensor, which collectively have been presented in \ref{Eq_Gal_rho} and \ref{Eq_Gal_p} respectively. Due to the complicated nature of the field equations it is not possible to solve them analytically, however numerical analysis can be performed. This explicitly shows that there are indeed scenarios in which late time acceleration can be achieved within this model, see \ref{Fig_EoS}. It turns out that for the ``Case A" in \ref{Fig_EoS} the re-defined Galileon field $\pi'$ must vary as $\sim t^{2}$ to the leading order, while the gauge field must grow as $\sim t^{3}$. On the other hand, the torsion tensor components must decrease in time with a similar power law behaviour. By modifying the power laws appropriately it is indeed possible to arrive at $\omega _{0}\sim -1$ (see e.g., ``Case B" and ``Case E" in \ref{Fig_EoS}), provided the coupling constant $\alpha$ is larger than certain bound. For example, if $\alpha>10^{-5}$, then we could obtain $\omega _{0} \leq -1/3$ (including $\omega _{0}\sim -1$) for various behaviours of the fields present in \ref{Eq_Gal_rho} and \ref{Eq_Gal_p} respectively. However for $\alpha \leq10^{-5}$, we could not obtain a scenario where late time acceleration is achieved (e.g., ``Case C" in \ref{Fig_EoS}). Thus for appropriate choices of the numerical value of the coupling constant and also for reasonable choices of the gauge field and the Galileon field one can indeed have late time accelerating solution of the universe, with torsion tensor playing the crucial role.

As evident from \ref{Fig_EoS}, the equation of state parameter changes with the redshift and hence is time dependent. It turns out that the above problem of time dependent equation of state parameter is equivalent to that of an interacting dark energy fluid, which has been extensively studied in \cite{Valiviita:2009nu,Said:2013jxa,Feng:1203,Tripathi:2016slv,Usmani:2008ce,Kopp:2018zxp,Crooks:2003pa,Olivares:2005tb,Duran:2010hi,Pan:2012ki}. Using Monte Carlo Markov Chain technique one can demonstrate that models involving interacting dark energy (or, time dependent equation of state parameter) has degeneracy between the matter density and the dark energy interaction rate. This being the prime reason that the associated constraints are much weaker. In particular, if we consider the two scenarios having equation of state parameter $\omega _{0} \sim -1$ (``Case B'' and ``Case E'' respectively), it is immediate that they show two different nature of time evolutions. In ``Case B'', the equation of state parameter evolves Logarithmically with redshift, while in ``Case E'', the evolution of the equation of state parameter is slower. Incidentally, such behaviours of the equation of state parameter, e.g., evolving Logarithmically, have already been addressed in \cite{Feng:1203,Tripathi:2016slv} taking into account of Supernova Type Ia data \cite{Astier:2005qq,Garnavich:1998th}, Baryon Acoustic Oscillation \cite{Delubac:2014aqe,Simon:2004tf} and measurements of Hubble parameter \cite{Samushia:2006fx,Farooq:2012ev}. It turns out that the numerical estimates of the best fit model obtained in \cite{Feng:1203,Tripathi:2016slv} is consistent with the numerical results derived in the present analysis. Thus such behaviours of the equation of state parameter appear naturally in various interacting dark energy models as well, which are consistent with the current cosmological measurements. This explicitly demonstrates the robustness and compatibility of our results with current observations.
\paragraph*{\bf Concluding Remarks} ---~~ We have shown that the Galileon model with spacetime torsion, introduced here, is able to predict the late time acceleration of the universe. It is important to point out that this is the first time spacetime torsion has been included in the covariant Galileon theory. Interestingly enough, the Galileon model itself can act as the source of torsion. Thus we need not have to invoke some exotic matter, e.g., spin fluid to generate spacetime torsion. We have also demonstrated that arbitrary choices for the spacetime torsion are unable to provide late time acceleration, one needs to consider a general situation compatible with homogeneity and isotropy of the universe in order to achieve the same. It turns out that in the general context as well there can be multiple situations with either power law or exponential time variation of the basic fields, where late time acceleration can be achieved, or, in other words the equation of state parameter at present epoch $\omega_{0}$ becomes less than or equal to $-1$. This in turn requires the coupling $\alpha$ to be larger than $10^{-5}$ in natural units. For smaller values, unless some fine-tuning is employed it is not possible to arrive at late time accelerating model of our universe. Moreover, all of these interesting features have been achieved in a natural and systematic manner by localizing the shift-symmetry and hence the theory remained ghost free even on the Riemann-Cartan manifold. 

The covariant Galileon model with torsion developed here opens up several directions of exploration in cosmology as well as in black hole physics. For example, it will be of significant interest to understand the cosmological perturbation theory, structure formation and more importantly the physics of Cosmic Microwave Background Radiation in the present context to assess whether the above model is compatible with the current Planck data. The issue of compatibility may gain further momentum as the Square Kilometer Array becomes operational. Therefore accurate theoretical estimation of various cosmological parameters in this model is important in view of current and upcoming experiments.  Besides, one can also look for non-trivial black hole solutions in this model with torsional hair and in particular how generation, propagation and finally detection of gravitational waves are affected in these models. These computations may find their relevance in the upcoming experiments e.g., LISA, which will probe the realm of gravitational wave astrophysics with significant detail. These estimations and computations we leave for the future.
\section*{Acknowledgements}

Research of SC is funded by the INSPIRE Faculty Fellowship (Reg. No. DST/INSPIRE/04/2018/000893) from Department of Science and Technology, Government of India.

\bibliography{References}

\bibliographystyle{./utphys1}
\end{document}